\documentclass[]{aastex7}
\usepackage{xcolor}

\newcommand\osdmin{\texttt{overshoot\_D\_min}} 
\begin{document}

\title{Exploring Mixing Thresholds in Asteroseismic Stellar Evolution Models} 

\correspondingauthor{Lynn Buchele}

\author[0000-0003-1666-4787]{Lynn Buchele} 
\affiliation{Heidelberg Institute for Theoretical Studies, Schloss-Wolfsbrunnenweg 35, 69118 Heidelberg, Germany}
\affiliation{Center for Astronomy (ZAH/LSW), Heidelberg University, Königstuhl 12, 69117 Heidelberg, Germany}
\email[show]{lynn.buchele@h-its.org}

\begin{abstract}
Inferences from observations clearly show that mixing in stars extends beyond the convective boundaries defined by mixing length theory. This triggered the proposal of a variety of prescriptions to include additional mixing in stellar models. These prescriptions typically introduce free parameters to set the extent of the additional mixing and may also introduce numerical parameters. In the case of exponential overshooting, one must decide the threshold at which the exponential decay of the mixing coefficient can be treated as zero. Using the MESA stellar evolution code, I explore the effect of varying this parameter on asteroseismic models of main-sequence stars with growing convective cores. From this, I conclude that \osdmin{} should be set to $10^{-2}$ cm$^2$/s or lower for these stars. The default value in MESA is four orders of magnitude higher than this recommendation, which results in discontinuous evolution. 
\end{abstract}

\section{Introduction} 
Convection in stars remains difficult to incorporate into stellar evolution calculations due to the huge span in length and time scales that must be considered. Mixing length theory (for a review, see, for example, \citealt{2023Galax..11...75J}) is commonly used in 1D stellar evolution calculations. However, it is clear from observations that some amount of mixing extends beyond the formal convective boundaries \citep[see,][and references therin]{2023Galax..11...56A}. A number of different approaches have been developed to account for this additional convective boundary mixing. These approaches often introduce free parameters to control the amount of additional mixing. The implementation of some prescriptions also introduces numerical parameters that may affect the resulting stellar structure. This work focuses on one numerical parameter in the commonly used exponential overshoot prescription. 

Exponential overshooting sets the mixing coefficient in the region outside the convective boundary to decay exponentially according to 
\citep{2000A&A...360..952H}
\begin{equation} 
D_{ov}(z) = D_0 \exp \left(\frac{-2z}{f H_p}\right), 
\label{equ:D_exp} 
\end{equation} 
where $z$ is the distance from the edge of the convection zone, $D_0$ is the value of the diffusion coefficient in the convective region near the boundary, $H_p$ is the pressure scale height, and $f$ is a free parameter that sets the extent of the overshooting. The exponential form means that the diffusion coefficient will never truly reach zero, and some threshold below which the mixing will be set to zero must be chosen. \citet{2000A&A...360..952H} defined this parameter as $D_{ov}^{\rm{limit}} = 10^{-2}$ cm$^2$/s. In recent versions of MESA \citep[][and references therin]{2023ApJS..265...15J}, this cutoff is set using the inlist parameter \osdmin{}, although in versions prior to r12778 this threshold was set using \texttt{D\_mix\_ov\_limit}. This work explores the effect of different values of \osdmin{} on the asteroseismic frequencies of stars with growing convective cores during part of their main-sequence evolution.

\section{Asteroseismic Frequency Evolution}  
Using MESA version 24.08.1, I modeled the main-sequence evolution of a 1.35$M_\odot$ star\footnote{All files necessary to reproduce this work are available at \url{https://zenodo.org/records/15827232}} using three different values of \osdmin{}, $10^2$ cm$^2$/s (the default value), $1$ cm$^2$/s, and $10^{-2}$ cm$^2$/s (the value adopted by \citealt{2000A&A...360..952H}). All other parameters were kept constant. Using the GYRE code \citep{Townsend2013}, I calculated the frequency of the radial mode with radial order\footnote{The results are insensitive to the choice of mode.} 19 for models with ages between 0.5 and 3 Gyr. I examined both the frequency evolution and the interpolation error. 
% To check if the frequency evolved smoothly, I examined both the frequency and its derivative with respect to time, computed using the finite difference method as implemented in \texttt{numpy.gradient}. 
% I also examined the interpolation error, since interpolating frequencies along a track is common in asteroseismic modeling. 
For this, I fit a cubic spline to the frequency evolution, excluding every fifth model, and calculated the interpolation error ($\epsilon_{\rm{interp}}$) of the excluded models as: 
\begin{equation}
\epsilon_{\rm{interp}} = \nu_{\rm{GYRE}} - \nu_{\rm{interp}}, 
\label{equ:interp_error}
\end{equation} 
where ($\nu_{\rm{GYRE}}$) and  ($\nu_{\rm{interp}}$) are calculated using GYRE and the cubic spline fit, respectively. Ideally,  ($\epsilon_{\rm{interp}}$) should be much lower than the uncertainty of observed frequencies, which for these stars observed by \emph{Kepler} is roughly 0.1$\mu$Hz. 

\begin{figure} 
\plotone{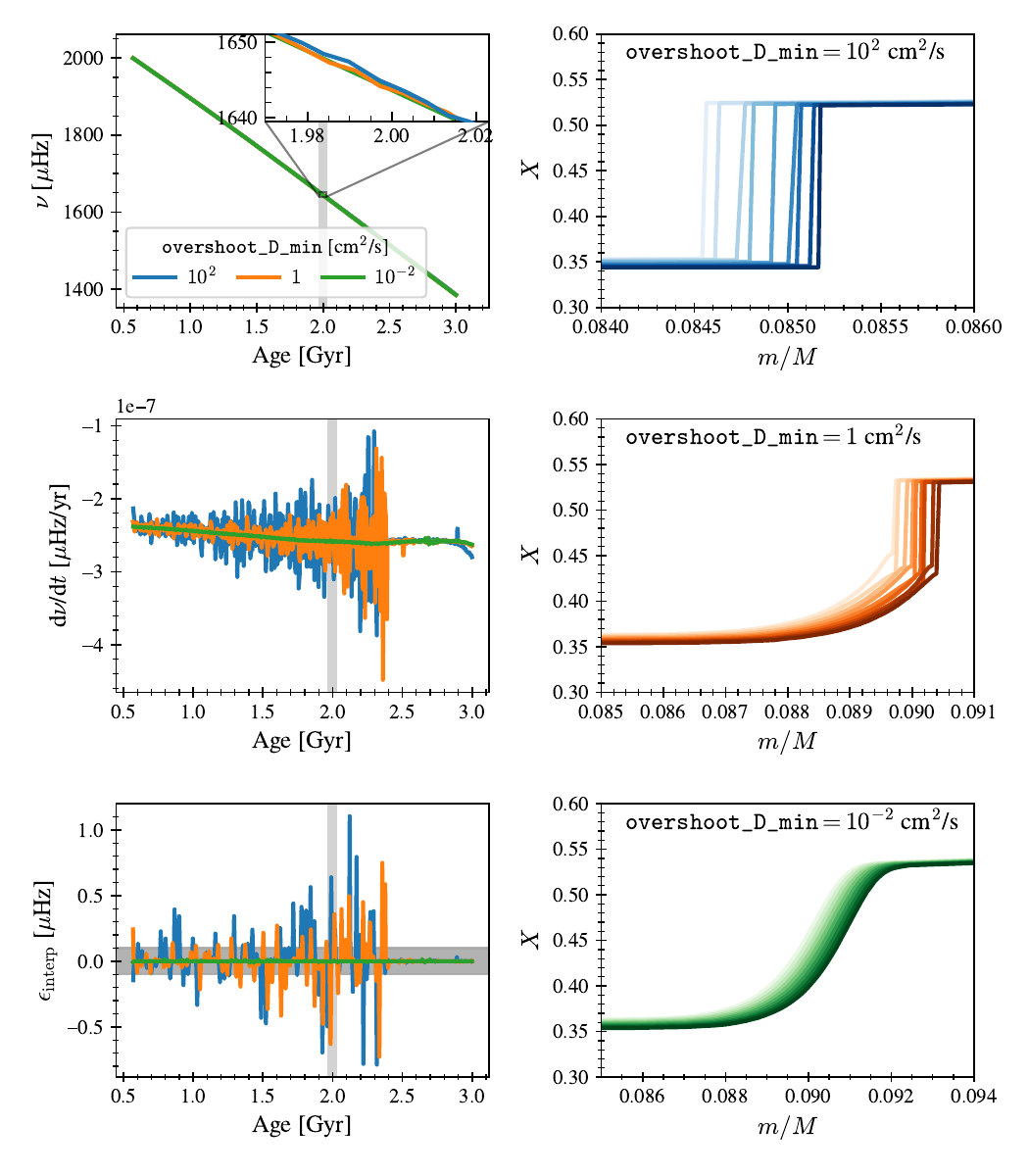} 
\caption{Left: 
The top panel shows the test mode frequency over time. The inset zooms in to show some differences in the frequency evolution. The middle panel shows the numerical time derivative of the test frequency. The bottom panel shows the interpolation error as defined in Equation~\ref{equ:interp_error}. The dark gray horizontal shading shows the 1$\sigma$ range typical of frequencies of stars observed by \emph{Kepler}.  In all three panels, the vertical light gray shading indicates the range of profiles plotted in the right column. \\
Right: Hydrogen abundance profiles around the overshoot boundary for selected profiles from each track. In all three panels, the lighter colored lines correspond to younger models.}

\label{fig:os_D_min} 
\end{figure}

I plot the frequency, frequency derivative (computed using \texttt{numpy.gradient}), and interpolation error for each of the three tracks in the left column of Figure~\ref{fig:os_D_min}. Zoomed out, the frequency evolution appears to be the same for all three tracks; however, the inset shows some deviation in the frequency evolution of each track. The derivative of the frequency is quite noisy for higher values of \osdmin{} over much of their evolution, and there is significant interpolation error when the convective core is growing, reaching values over 1$\mu$Hz. These reduce significantly when the convective core begins to recede at around 2.4 Gyr. 

The right column of Figure~\ref{fig:os_D_min} shows the hydrogen mass fraction profile around the overshoot boundary for several successive models of each track. When \osdmin{} =  $10^2$ cm$^2$/s, the abundance profiles show a sharp step at the end of the overshoot region. While $D_{ov}$ does decrease in the overshoot region, it is still high enough to efficiently mix the stellar material even at the boundary where $D_{ov}$ = \osdmin{}. The location of this sharp change depends on where MESA places mesh points, which can change between successive models, and thus, the edge of the overshooting region does not evolve smoothly with time, as shown by a difference in the slope of the abundance profile between different models. 

When \osdmin{} is lowered to $1$ cm$^2$/s, the hydrogen abundance changes smoothly until the threshold set by \osdmin{} is reached. At this point, the mixing region is truncated, and the composition changes sharply. As with the default case, the point where the overshooting ends does not change smoothly between different models. 

When \osdmin{} is set to $10^{-2}$ cm$^2$/s, the hydrogen abundance profile is smooth throughout the entire overshoot region and evolves smoothly from one model to the next. This smooth evolution of the composition profile creates the smoother frequency evolution seen in the left column.

The frequencies of the star are most sensitive to the truncation of the overshooting zone during the growth of the convective core. This is because as the convective core grows, it mixes fresh hydrogen into the core, changing the abundance profile in the burning regions of the star. However, when the core recedes, the truncation of the overshooting only affects the abundance profile of future models outside the burning regions, significantly reducing the numerical noise in the frequency evolution.

\section{Conclusions} 
Based on the tests described above, I recommend that users set \osdmin{} (or the equivalent parameter in other codes) to $10^{-2}$ cm$^2$/s or lower when modeling main-sequence stars with growing convective cores using exponential overshooting. This is particularly important for models that will be used for asteroseismology, as abrupt truncation of the exponential overshooting leads to noisy evolution of frequencies, resulting in significant interpolation errors.  This parameter may also affect models of other masses or evolutionary stages where exponential overshoot is used, such as massive main-sequence stars, more evolved stars exhibiting hydrogen shell burning, or stars with more advanced burning (helium and beyond). Beyond the physical changes in the models, the smoother mixing profiles resulting from a lower value of \osdmin{} should also result in models that are more numerically robust. Depending on the results of a broader set of tests, it may be advisable to change the default value in MESA to the originally proposed value of 10$^{-2}$ cm$^2$/s.

%% Please use the acknowledgment and contribution environments. This will 
%% be anonymized when the "anonymous" style option is used. 
\begin{acknowledgments}
I wish to thank Earl Bellinger and Ebraheem Farag for providing comments on an earlier draft.
\end{acknowledgments}

% \begin{contribution}
%%This section gives authors the space to recognize author contributions. The text inside this environment is NOT counted towards the total word quanta. At a minimum, manuscripts are expected to include this text:

% Contribution statement here
%% But authors are expected to provide more specific details, e.g. 
%%
%%SC was responsible for writing and submitting the manuscript.
%%WWM came up with the initial research concept and edited the manuscript.
%%OTS obtained the funding and edited the manuscript.
%%EBF provided the formal analysis and validation. He also edited the manuscript.
%%GEH Supervised the undergraduates, wrote the software and administers the project github and Zenodo repositories.
%%
%% Authors can use the Contributor Role Taxonomy (CRediT) at
%% for ideas on how write a good statement tailored to their needs.

% \end{contribution}

% \facilities{HST(STIS), Swift(XRT and UVOT), AAVSO, CTIO:1.3m, CTIO:1.5m, CXO}

% \software{astropy \citep{2013A&A...558A..33A,2018AJ....156..123A,2022ApJ...935..167A} }

\bibliography{os_D_min}{}
\bibliographystyle{aasjournalv7}


\end{document}